# Probing the electron-phonon coupling in $MgB_2$ through magnetoresistance measurements in neutron irradiated thin films

M. Monni[1,2], I. Pallecchi[2], C. Ferdeghini[2], V. Ferrando[2], A. Floris[3], E. Galleani d'Agliano[2], E. Lehmann[4], I. Sheikin[5], C. Tarantini[2], X.X. Xi[6], S. Massidda[1], M. Putti[2]

[1] CNR-INFM-SLACS and Department of Physics University of Cagliari, Cittadella Universitaria, I-09124 Monserrato (CA), Italy

[2] CNR-INFM-LAMIA and Dipartimento di Fisica, Università di Genova, Via Dodecaneso 33, 16146 Genova, Italy

[3] Institut für Theoretische Physik, Freie Universität Berlin, Arnimallee 14, D-14195 Berlin, Germany

[4] Paul Scherrer Institut, CH-5232 Villigen, Switzerland

[5] GHMFL, MPI-FKF/CNRS, 25 Avenue des Martyrs, BP 166, 38042 Grenoble Cedex 9, France

[6] The Pennsylvania State University, University Park, PA 16802, USA

**Abstract**

We report magnetoresistance (MR) measurements on $MgB_2$ and the corresponding full account from *ab-initio* calculations; we suggest that this combination can be a useful tool to probe electron-phonon coupling. We obtain good quantitative agreement between high field measurements on neutron irradiated epitaxial thin films and calculations within Bloch-Boltzmann transport theory over a wide range of magnetic fields (0-28 T) and temperatures (40-300 K), and as a function of the field orientation. The crossovers between in-plane and out-of-plane MR, experimentally observed as a function of either disorder or temperature are well reproduced indicating that disorder and interaction with phonons strongly affect the scattering rate of σ-carriers.

It is widely accepted that electron-phonon is the pairing mechanism responsible for the occurrence of superconductivity around 40K in magnesium diboride; it is therefore very important to study the phonon dispersion and the electron-phonon coupling (EPC), which account for such a high value of critical temperature. Several theoretical studies have been carried out in order to obtain the Eliashberg electron-phonon spectral functions $\alpha^2(\omega)F(\omega)$ [1,2,3,4], either global or band-resolved; there is presently no general agreement among the different calculations. In parallel, different experimental techniques have been employed for the investigation of lattice dynamics and EPC. In particular, EPC was monitored by electron tunneling spectroscopy [5], optical conductivity [6], photoexcited quasiparticle dynamics [7], magnetic field penetration depth [8] and de Haas-van Alphen effect [9]. The electron-phonon interaction also determines transport properties, such as the electrical dc resistivity $\rho$; unfortunately, $\rho$ is affected by poor intergrain connectivity [10] and cannot therefore provide reliable information. On the other hand, the combined dependence of the resistivity on temperature and magnetic field provides a far more reliable tool to get information on intrinsic transport in $MgB_2$ [11]. Being defined as a relative variation, magnetoresistance (MR) is not affected by uncertainties related to unknown geometrical factors. Due to its sensitivity to disorder, experimental MR varies in a broad range of values, [12,13,14,15] up to 136% at 18 T in the cleanest samples [14]. MR has already been used to estimate the ratio of scattering times in $\pi$ and $\sigma$ bands in polycrystalline [12] and epitaxial thin films [13].

Due to the anisotropic character of magnetotransport, measurements in epitaxial thin films exhibit a complex phenomenology. The in plane MR can be higher (lower) than the out of plane MR depending on disorder [10,13] and temperature [14]. Moreover, by varying the angle between the magnetic field and the crystallographic axes MR oscillates showing different shapes as a function of temperature and magnetic field [14]. This indicates that the combination of Fermi surface (FS) topology and of a large anisotropic (over different bands) EPC determines the resulting MR in a non-trivial way: once the FS topology is taken into account MR becomes a very suitable and precious tool to gain information on the EPC selectively in the two kinds of bands. On the other hand, it is clear that reliable hints on EPC can be extracted from MR only if scattering rates by impurities are not much larger than those by phonons.

In this letter, we combine an experimental investigation of MR on undamaged and neutron irradiated epitaxial thin films with a full account of results based of *ab-initio* electronic structure calculations, within Bloch-Boltzmann transport theory. By considering the sample dependent impurity scattering rates as free parameters, we could obtain a very good agreement with experimental data over the whole magnetic field and temperature range. At low temperature, the dependence of isothermal MR on the field orientation, which mirrors the balance between $\sigma$ and $\pi$ contributions mediated by the FS topology, is also very well accounted for by the calculations.

High quality epitaxial magnesium diboride films, with low residual resistivity $\rho_0$ and high critical temperature (~41K) were grown on 4H-SiC by Hybrid Physical Chemical Vapour Deposition [16]. The films were deposited on the same run. Two of them, IRR15 and IRR25, were irradiated at the spallation neutron source SINQ of the Paul Scherrer Institute in Villigen [17]. The main properties of the films are summarized in Table I. Because MR strongly decreases with increasing resistivity, only weakly irradiated samples were investigated: after irradiation $T_c$ decreases only from 41.4 to 38.3 K and the residual resistivity increases from 0.55 to 8.6 $\mu\Omega$cm.

High field MR measurements were performed at the High Magnetic Field Laboratory in Grenoble. Measurements were carried out at fixed temperatures between 42 K and room temperature and in magnetic fields up to about 28 T with the magnetic field applied parallel and perpendicular to the film surface, always perpendicular to the current.

We calculate MR by solving numerically the Boltzmann transport equation in the relaxation time approximation [18]; in this framework the dc electrical conductivity tensor in uniform magnetic field is given by:

$$\sigma(\mathbf{B},T) = \int \sigma(\varepsilon,\mathbf{B},T)\left(-\frac{\partial f(\varepsilon,T)}{\partial \varepsilon}\right)d\varepsilon, \text{ where} \tag{1}$$

$$\sigma(\varepsilon,\mathbf{B},T) = e^2 \sum_{i=\sigma,\pi} \int_{BZ} \frac{d\mathbf{k}}{4\pi^3} \tau^i(\mathbf{k},\varepsilon,T)\mathbf{v}_i(\mathbf{k})\bar{\mathbf{v}}_i(\mathbf{k},\mathbf{B})\delta(\varepsilon_i(\mathbf{k})-\varepsilon) \tag{1a}$$

where $\tau^i(\mathbf{k})$ is the scattering times of the $i$-th band and $f(\varepsilon,T)$ is the Fermi function; $\mathbf{v}_i(\mathbf{k})$ is the Fermi velocity and $\bar{\mathbf{v}}_i(\mathbf{k},\mathbf{B})$ is its weighted average over the past history of the electronic wave-packet, $\bar{\mathbf{v}}_i(\mathbf{k}) = \int_{-\infty}^{0} \frac{dt}{\tau^i(\tilde{\mathbf{k}}_i(t))} e^{t/\tau^i(\tilde{\mathbf{k}}_i(t))} \mathbf{v}_i(\tilde{\mathbf{k}}_i(t))$. The magnetic field $\mathbf{B}$ determines the time evolution of $\tilde{\mathbf{k}}_i(t)$, which samples the FS with the cyclotron frequency $\omega_c$; when $\omega_c \tau$ is sufficiently large the relative resistivity change is significant, under geometrical conditions dictated by the FS topology and by the orientation of $\mathbf{B}$. We solve the semiclassical equation of motion $\frac{d\tilde{\mathbf{k}}_i(t)}{dt} = \frac{-e}{\hbar} \mathbf{v}_i(\tilde{\mathbf{k}}_i(t)) \times \mathbf{B}$ at each $\mathbf{k}$ by numerical integration, with the initial condition $\tilde{\mathbf{k}}_i(0) = \mathbf{k}$.

Within the framework of the relaxation time approximation the electronic scattering times are assumed to be dependent of $\mathbf{k}$ just through the corresponding energy, within each band type: $\tau^i(\mathbf{k}) = \tau^i(\varepsilon(\mathbf{k}))$. More specifically, $\tau^i$ is a function of both energy and temperature: $\frac{1}{\tau^i(\varepsilon,T)} = \frac{1}{\tau^i_{imp}} + \frac{1}{\tau^i_{ph}(\varepsilon,T)}$. Here $\tau^i_{imp}$ is the sample-dependent lifetime from the impurity scattering, taken to be independent of both energy and temperature, and $\tau^i_{ph}(\varepsilon,T)$ is the electronic life time from scattering with phonons, which depends on both energy and temperature. $\tau^i_{ph}(\varepsilon,T)$ can be obtained from the transport Eliashberg functions $\alpha^2_{Tr}F_i(\omega) = \sum_j \alpha^2_{Tr}F_{ij}(\omega)$ as [19]:

$$\frac{1}{\tau^i_{ph}(\varepsilon,T)} = \left(1 + e^{-\beta\varepsilon}\right) \int_{-\omega_{max}}^{\omega_{max}} \frac{2\pi \alpha^2_{Tr}F_i(\omega)}{\left(e^{\beta\omega}-1\right)\left(1+e^{-\beta(\varepsilon+\omega)}\right)} d\omega \qquad (2)$$

$\alpha^2_{Tr}F_i(\omega)$ are calculated using density functional perturbation theory [20]. This is done with a careful Brillouin zone integration, considering a 24×24×24 and a 8×8×8 mesh for $\mathbf{k}$ and $\mathbf{q}$ points (electron and phonon momenta) respectively. These functions correspond to the values of the EPC $\lambda_{\sigma,Tr} = 0.93$ and $\lambda_{\pi,Tr} = 0.43$. The major difference between the transport and superconducting

Eliashberg functions in MgB$_2$ is found in the role played in the former by the $E_{2g}$ phonon modes, strongly coupled to $\sigma$ bands: we have a reduced importance for $q$ along the $\Gamma$-A line of the Brillouin zone, corresponding to zero in-plane momentum transfer, and an amplified role at large in-plane $q$. In the limit of a negligible energy variation of physical quantities around $E_F$, we have $\left(\tau_{ph}^i\right)^{-1} = \int d\omega \sum_j \Gamma^{ij}(\omega,T)$, where $\Gamma^{ij}(\omega,T) = \pi\beta\,\omega\alpha_{Tr}^2 F_{ij}(\omega)/\sinh^2(\beta\omega/2)$ represents the band-resolved frequency-dependent phononic contribution to the scattering rate, that we only use for graphical visualization [21]. In Fig. 1 we plot, at two selected temperatures, $\Gamma^{ij}(\omega,T)$. While below about 100 K the contribution is negligible, with increasing temperature it progressively grows and this is particularly true for the more coupled $\sigma$ bands. This implies that the temperature dependence of the scattering rates is strongly influenced by the spread of phonon frequencies, whose influence can therefore be monitored through the temperature dependence of MR.

In Fig. 2 we present the results of selected isothermal MR measurements. We plot normal state MR data, measured with **B** either parallel or perpendicular to the *ab* planes, as a function of the squared magnetic field, for different samples at the lowest temperature (panels in the left column) as well as for the unirradiated samples at different temperatures (first panel in the left column and panels in the right column). The behaviour of MR as a function of B$^2$ changes and its magnitude decreases with increasing *T* or disorder, as it is crucially sensitive to the electronic lifetime $\tau$. In the unirradiated sample at the lowest temperature T~43 K, the configuration with **B**$\perp$*ab* results in the largest MR of nearly 70% at 28T. In sample IRR15 the MR is lower and the curves for **B**∥*ab* and **B**$\perp$*ab* tend to merge (second panel of the left column). In the case of IRR25, the MR is less than 3% and for **B**∥*ab* is even slightly larger than that for **B**$\perp$*ab* (third panel of the left column). A similar crossover occurs for the unirradiated sample with increasing temperature. This behaviour can be rationalized by considering the coexistence in MgB$_2$ of $\sigma$ and $\pi$ bands with two-dimensional (2D) and three-dimensional (3D) character, respectively: the $\sigma$ electrons under the influence of magnetic field **B**∥*ab* move along open **k**-space trajectories, with nearly constant Fermi velocities $\bar{\mathbf{v}}_i(\mathbf{k}) \approx \mathbf{v}_i(\mathbf{k})$, reducing to straight lines in an exactly 2D case; the in-plane conductivity, then, would be independent of the magnetic field. The crossover is then justified by assuming cleaner 2D σ bands at low *T* in the unirradiated samples. The stronger EPC for σ bands decreases the corresponding $\tau_\sigma$ at high *T*; the same effect is obtained by disorder. The contribution of π bands eventually yields a larger MR for **B**||*ab*. The π contribution to MR is in fact intrinsically isotropic, but due to the larger mobility of π carriers in the direction ⊥*ab* than in the direction ||*ab*, MR for **B**||*ab* is larger than for **B**⊥*ab* [10].

This interpretation is quantitatively supported by first principles calculations, compared with experimental results in Fig. 2. With only two unavoidable free parameters for each sample (the impurity scattering rate $\tau_{imp}^i$ for the two bands) the behaviour of MR as a function of *B* and T is well described for all samples. At low *T*, when MR is dominated by scattering with impurity, the curves are well reproduced; then, increasing the temperature, switching on the electron-phonon contribution, the whole set of MR curves is also remarkably well described; in particular, the crossovers between the two field orientations is traced, both in terms of temperature and sample disorder. In order to highlight this behaviour, Fig. 3 shows the MR at the highest field (28 Tesla) as a function of temperature. It is clearly seen that the crossover between perpendicular and parallel MR is fairly well reproduced by taking into account the temperature dependence of $\tau_{ph}^i(\varepsilon,\mathrm{T})$ determined by the Eliashberg functions. The agreement between theory and experiment is striking, both in terms of values (only at the highest T the agreement deteriorates, but the MR value itself is very small) and crossing temperature or lack of it in the heavily irradiated sample.

The values of the free parameters $\tau_{imp}^i$ are reported in Table 1 for each sample. We see that the unirradiated sample has cleaner σ bands ($\tau_{imp}^\sigma / \tau_{imp}^\pi$ ~5.5); irradiation reduces both lifetimes, resulting into almost equal damages in the two types of bands. Indeed in IRR25 sample $\tau_{imp}^\sigma / \tau_{imp}^\pi$ ~1.5. IRR15 sample have been irradiated very lightly, yet sufficient to decrease $\tau_{imp}^\sigma$ by roughly a factor of two; the value of $\tau^{\pi}_{imp}$ is slightly larger than the UNIRR sample, probably as a consequence of the uncontrolled difference between the different pristine samples.

In Fig. 4 we plot $\tau^{\pi}(T)$ and $\tau^{\sigma}(T)$, defined by $1/\tau^{i}(T) = 1/\tau^{i}_{imp} + \int \sum_{j} \Gamma^{ij}(\omega,T) d\omega$; for sake of clarity only the curves of the samples UNIRR and IRR25 are reported. The sample UNIRR has cleaner σ bands; since irradiation damages almost evenly the two types of bands, the sample IRR25 has lower and nearly comparable scattering times. With increasing temperature the electron-phonon interaction lowers more efficiently the lifetime of the more coupled σ bands; the scattering rates decrease and cross at about 180-190 K in both the samples. As expected, with increasing temperature, $\tau^{\sigma}$ and $\tau^{\pi}$ curves pass from a sample dependent regime dominated by impurity scattering to an intrinsic regime dominated by phonon scattering; in the latter, the scattering times of different samples tend to merge and $\tau^{\sigma}/\tau^{\pi}$ becomes less than one, indicating that transport is eventually dominated by phonon scattering of the π carriers.

The very good agreement of the *T*-dependent MR actually raises the question of the use of this technique, accompanied by theoretical calculations, to monitor the reliability of calculated EPC functions. This may prove particularly useful in MgB$_2$, where the calculation of $\alpha^2 F$ has proven to be very difficult.

In summary, we reported the first complete account of magnetoresistance in both irradiated and undamaged MgB$_2$ as a function of temperature, magnetic field magnitude and orientation. The experimental data are compared with results of density functional based calculations within Bloch-Boltzmann theory. The excellent agreement demonstrates the reliability of the calculated transport Eliashberg functions and allows extracting the phonon scattering rates in the two bands as a function of temperature. This success is a clear hint that this method can validly complement other experimental techniques in probing EPC in superconducting materials.

This work is supported by MIUR under the projects PRIN2006021741 and PON CyberSar, by INFM-CNR through a supercomputing grant at Cineca Bologna, Italy, by NSF under Grant No.

DMR-0306746 and by ONR under Grant No. N00014-00-1-0294 and by the European Commission from the 6th framework programme "Transnational Access - Specific Support Action", contract RITA-CT-2003-505474.

**Figure captions**

FIG. 1. (color online) $\Gamma^{ij}(\omega,T) = \pi\beta\,\omega\,\alpha_{Tr}^{2}F_{ij}(\omega)/\sinh^{2}(\beta\omega/2)$ vs ω evaluated at T = 160 K (red lines) and 300 K( black lines).

FIG. 2. (color online): MR curves for **B** either parallel (open red circles: experimental data; blue triangles: calculation) or perpendicular (filled black circles: experimental data; green squares: calculation) to the *ab* planes plotted as a function of B$^2$. Data for different samples with increasing disorder at the lowest temperature (panels of the left column, from top to bottom UNIRR at T=43

K, IRR15 at 50 K and IRR25 at 42 K) and for the unirradiated sample at different temperatures (panels of the right column, from top to bottom UNIRR at T=100K, at 160K and at 300K).

FIG. 3. Experimental (filled symbols) and theoretical (open symbols) MR for B = 28 T parallel (left triangle) and perpendicular (triangle up) to the *ab* planes. Results are plotted as a function of the temperature for different samples (lines are guide to the eyes).

FIG. 4. Temperature dependence of band-resolved electronic lifetimes in UNIRR and IRR25 samples.

**Table caption**

**Table I**: Main properties of samples: neutron fluence, critical temperature (defined as 50% of residual resistivity), residual resistivities $\rho_0$ (calculated at *42 K*), RRR= $\rho(300)/\rho(42)$, impurity scattering times as extracted from the fitting procedure.

| Sample | Neutron Fluence (cm$^{-2}$) | $T_c$ (K) | $\rho_0$ ($\mu\Omega \cdot cm$) | RRR | $\tau_{imp}^{\pi}$ ($10^{-14}$ s) | $\tau_{imp}^{\sigma}$ ($10^{-14}$ s) |
|---|---|---|---|---|---|---|
| IRR 0 | -- | 41.4 | 0.55 | 9 | 4.0 | 22.10 |
| IRR 15 | 4.1·10$^{16}$ | 40.3 | 2.7 | 7 | 5.5 | 11.15 |
| IRR 25 | 4.1·10$^{17}$ | 38.3 | 8.6 | 2.4 | 1.8 | 2.65 |

**Table 1.**

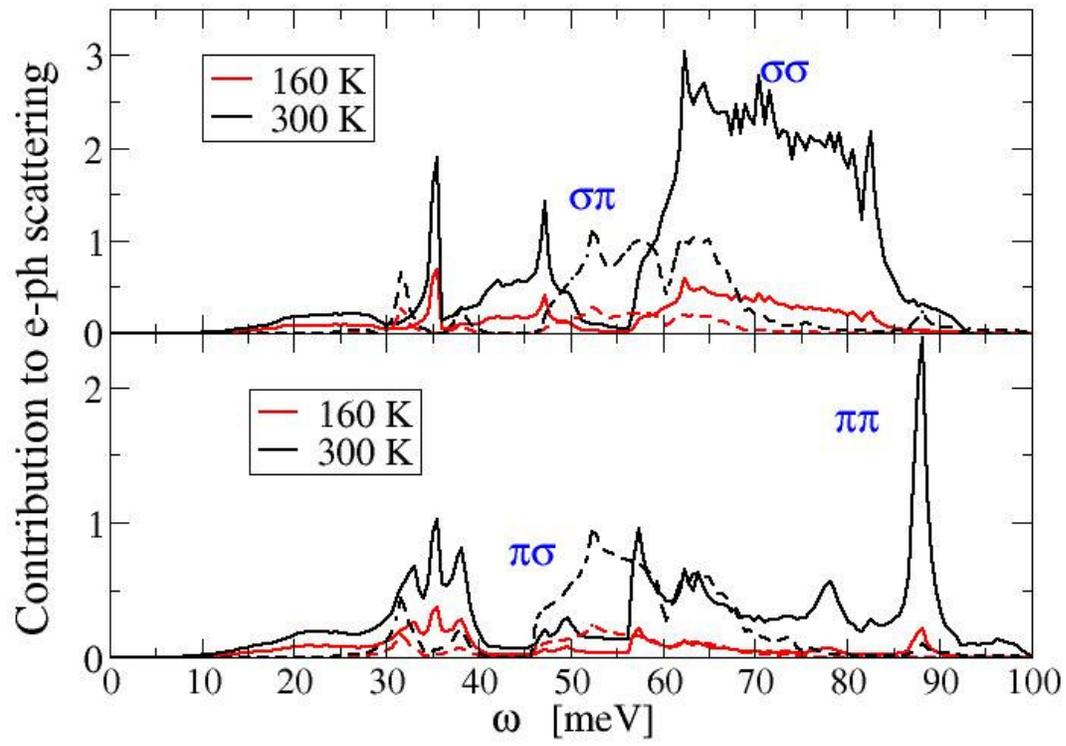

**Figure 1.**

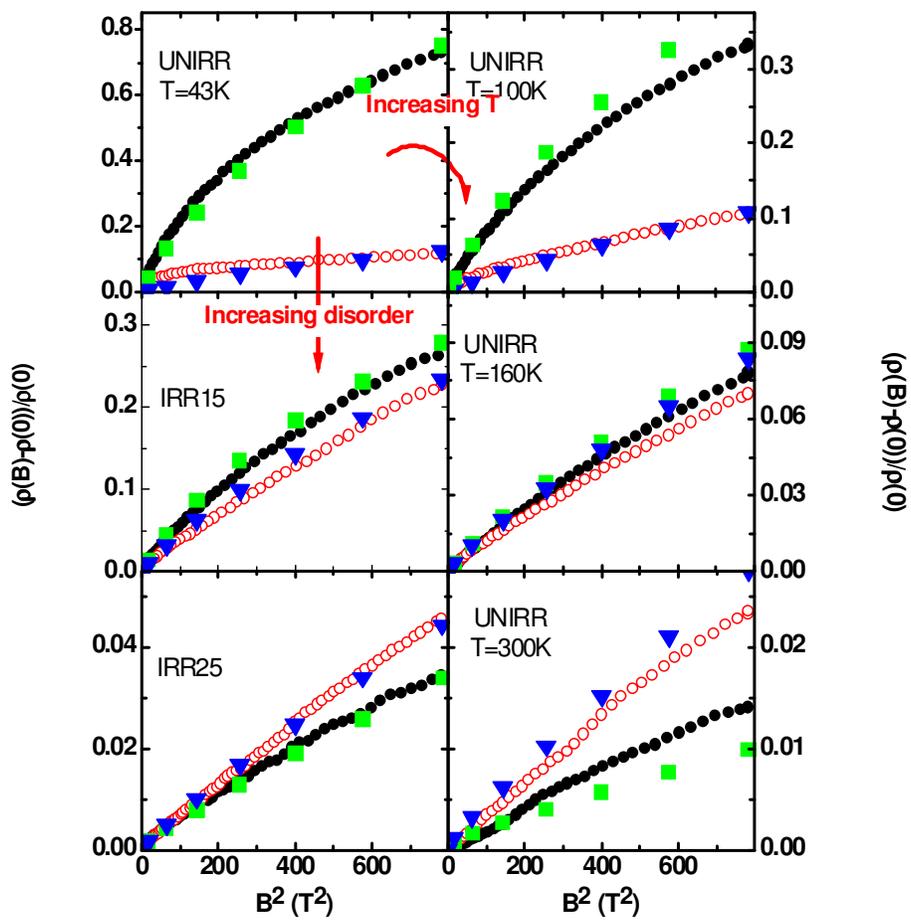

Figure 2.

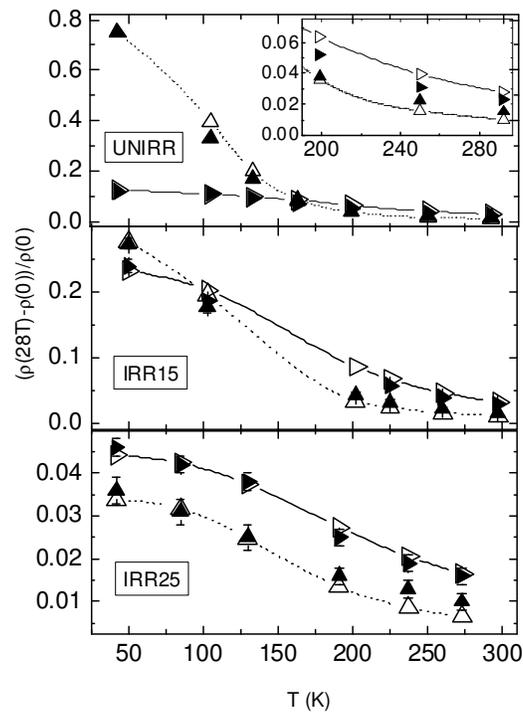

Figure 3.

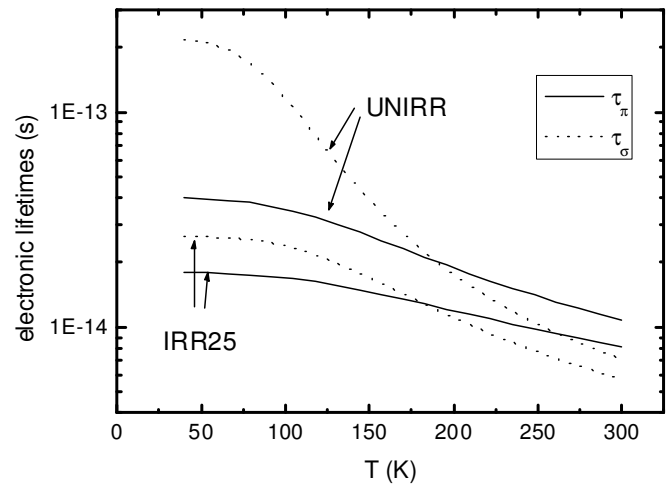

Figure 4.